\documentclass[twocolumn,prc,nofootinbib,showpacs,epsfig]{revtex4}
\usepackage{graphicx}

\newcommand{\ra}{\rightarrow }
\def\Journal#1#2#3#4{{#1} {\bf #2}, #3 (#4)}
\def\NCI{\em Nuovo Cimento}

\def\NPA{{\em Nucl. Phys.} A}

\def\PLB{{\em Phys. Lett.} B}

\def\JPG{\em Journal of Physics G}
 
\def\PRL{\em Phys. Rev. Lett.} 
\def\PR{\em Phys. Rev.} 
 
\def\PRC{{\em Phys. Rev.} C} 
\def\ZPC{{\em Z. Phys.} C} 
\def\PTP{{\em Prog. Theor. Phys.}}

\def\PAN{{\em Physics of Atomic Nuclei}}

\def\ANDT{\em Atomic and Nuclear Data Tables}
\newcommand{\be}{\begin{equation}}
\newcommand{\ee}{\end{equation}}
\def\bea{\begin{eqnarray}} 
\def\eea{\end{eqnarray}} 

\newcommand{\obb}{0\mbox{$\nu\beta\beta$ - decay} } 
\newcommand{\zbb}{2\mbox{$\nu\beta\beta$ - decay} }

\newcommand{\moeh}{\mbox{$^{100}$Mo }}

\newcommand{\tinhz}{\mbox{$^{112}$Sn }}
\newcommand{\cdhz}{\mbox{$^{112}$Cd }}

\newcommand{\tinhv}{\mbox{$^{124}$Sn }}
\newcommand{\tehv}{\mbox{$^{124}$Te }}
 
\newcommand{\ndhf}{\mbox{$^{150}$Nd} }

\newcommand{\bpbp}{\mbox{$\beta^+\beta^+$} }
\newcommand{\ecec}{\mbox{$EC/EC$} }

\newcommand{\bec}{\mbox{$\beta^+/EC$} }
\newcommand{\ton}{\mbox{$T_{1/2}$}}

\newcommand{\bnel}{\mbox{$\bar{\nu}_e$} }

\newcommand{\nel}{\mbox{$\nu_e$} }

\newcommand{\neu}{neutrino }

\newcommand{\ema}{\mbox{$\langle m_{\nu_e} \rangle $}} 
\newcommand{\iso}[2]{{\ensuremath{{}^{#2}\mathrm{#1}}}}

\begin{document}
\flushbottom
\title{A search for double beta decays of tin isotopes with enhanced sensitivity}
\author{J. Dawson$^{a,c}$, D. Degering$^{b}$, M. K\"ohler$^{b}$, R. Ramaswamy$^{a}$,C. Reeve$^{a}$, J. R. Wilson$^{a,d}$, K.
Zuber$^{a,e}$}
\affiliation{
$^{a}$Dept.~of Physics and Astronomy, University of Sussex, Falmer, Brighton BN1 9QH, UK\\
$^{b}$ Verein f\"ur Kernverfahrenstechnik und Analytik Rossendorf eV, Postfach 510119, 01314 Dresden, Germany\\
$^{c}$Laboratoire APC, B\^{a}timent Condorcet, 10, rue Alice Domon et L\'{e}onie Duquet, 75205 Paris Cedex 13, FRANCE\\
$^{d}$Dept.~of Physics, University of Oxford, Denys Wilkinson Building, Keble Road, Oxford, OX1 3RH, UK\\
$^{e}$Institut fuer Kern und Teilchenphysik, Technische Universitaet Dresden, Zellescher Weg 19, 01069 Dresden, Germany}
\begin{abstract}
A search for the various double beta decay modes of $^{124}$Sn and $^{112}$Sn has been performed on 75\,kg$\cdot$days of data. New half-life limits for excited states in $^{124}$Sn have been obtained including a lower limit for decay into the first excited 2$^+$ state of $^{124}$Te of $T_{1/2}>0.87~\times~10^{20}~\mbox{yrs}~(90\% CL) $ and into the first excited 0$^+$ state of $T_{1/2}>1.08~\times~10^{20}~\mbox{yrs}~(90\% CL)$. Ground state and excited state transitions of $^{112}$Sn have also been experimentally explored. A limit for the $2\nu EC/EC$ of $T_{1/2}>1.8~\times~10^{19}~\mbox{yrs}~(90\% CL)$ is obtained. The non-observation of de-excitation gammas from the 0$^+$ at 1888.5\,keV results in a lower half-life limit on the $0\nu EC/EC$ decay of $^{112}$Sn of  $T_{1/2}>0.8~\times~10^{19}~\mbox{yrs}~(90\% CL)$,  despite a possible resonant enhancement of the decay rate due to degenerated states.
\end{abstract}


\maketitle
\section{Introduction}
The Standard Model of Particle Physics has been extremely successful but is not thought to be the final theory.
The quest for physics beyond the Standard Model is gathering pace, with searches performed at accelerators such as Tevatron, HERA and soon at the LHC.  There are also non-accelerator experiments, such as the search for the rare process known as Neutrinoless Double Beta Decay.


The classical Double Beta Decay is allowed by the Standard Model.  In this reaction two neutrons in the same nucleus decay simultaneously to emit two anti-neutrinos and two electrons.  This possible reaction was first discussed by M. Goeppert-Mayer\cite{goe35} in the form of
\be
(Z,A) \ra (Z+2,A) + 2 e^- + 2 \bar{\nu_e} \quad (\zbb).
\ee
As this is a higher order process, practically to observe this requires an isotope in which single beta decay is forbidden or strongly suppressed.  The energy spectrum from the two electrons has similar features to that from single beta decay, a continous spectrum ending at a well defined end-point determined from the Q-value of the reaction. 

The classic papers of Majorana \cite{maj37}, in which he discussed a two-component neutrino, led Racah \cite{rac37} and Furry to first discuss a neutrinoless mode in the form of \cite{fur39}
\be
\label{proc0nu}
(Z,A) \ra (Z+2,A) + 2 e^-  \quad (\obb).
\ee
The signature of this mode is very different to the classical case.  Here the summed energy of the emitted two electrons is exactly equal to the Q-value of the reaction.  The experimental concept therefore is to observe an isotope capable of decaying via double beta decay and to search for a peak at the endpoint of the spectrum.

This process clearly violates total lepton number conservation by two units and so is forbidden by the Standard Model.  This is in contrast to neutrino oscillations which violate individual flavour lepton number but conserve total lepton number.  For this decay to occur neutrinos must be massive Majorana particles, however any $\Delta L =2 $ process can contribute to the decay.  The most interesting mode is the light Majorana neutrino exchange. Here the measured quantity is known as the effective Majorana \neu mass which is given by

\be 
\label{eq:ema}\ema = \mid\sum_{i} U_{ei}^2 m_i\mid =  \mid\sum_{i} \mid U_{ei} \mid^2 e^{2i\alpha_i} m_i \mid 
\ee 
which for the case of CP-invariance, $e^{2i\alpha_i} = 0, \pi$, is 
\be
\ema = \mid  m_1  U_{e1}^2 \pm m_2  U_{e2}^2 \pm m_3 U_{e3}^2  \mid
\ee 

For double beta decay the decay rate and thus inverse half-life is strongly dependent on the available Q-value.  In the classic case, $2\nu\beta\beta $-decay, the rate is proportional to Q$^{11}$ whilst the \obb\ mode scales with Q$^5$. For this reason, only high Q-value decays (typically above 2\,MeV) are currently worth studying.  This criterion restricts the 35 candidate isotopes to 11. 


There are also double beta decay modes (both \zbb and $0\nu\beta\beta$-decay) in which gamma rays are emitted in addition to the two electrons, thus giving these modes distinct signatures.  In principle, a finely segmented detector could observe the two electrons in one segment and simultaneously the emitted gamma ray in another segment.  Since the energies of the electrons and gamma ray are known this gives a strong experimental signature which is distinct from the background.  Studying these modes may shed some light on new physics and provide some new information for nuclear matrix element calculations.  Despite their reduced Q-values there are benefits in studying these modes.

These modes come from two sources, double beta decays to excited 0$^+$ and 2$^+$-states of the daughter and double positron decays in which positrons are emitted instead of electrons.  Observations of \zbb\ transitions into the first excited 0$^+$-state for \moeh and \ndhf have already been reported \cite{bar95,bar04,arn07,bar07}.


For the double positron decays, combinations of electron capture (EC) and positron emission can occur:
\bea
(Z,A) \ra (Z-2,A) + 2 e^+ + (2 \nel) \quad \bpbp\\
e^- + (Z,A) \ra (Z-2,A) + e^+ + (2 \nel) \quad \bec\\ 
2 e^- + (Z,A) \ra (Z-2,A) + (2 \nel)  \quad \ecec
\eea
In particular, the \bec mode shows an enhanced sensitivity to right handed weak currents\cite{hir94}. For each positron there is the possibility of also observing one or both of the 511\,keV photons.  These modes therefore can provide an extremely clean signature of up to five coincident energy deposits.  However, for each generated positron the available Q-value is reduced by $2m_{e}c^{2}$, which leads to much smaller decay rates than in comparable 0\mbox{$\nu\beta\beta$ - decay}. For $\beta^+\beta^+$-decay to occur, the Q-values must be at least 2048\,keV. Only six candidate isotopes are known to have such a high Q-value.

The full Q-value is available only in the \ecec mode but is difficult to detect. In a neutrinoless \ecec to the ground state of the daughter, a monoenergetic internal bremsstrahlung must be emitted and requires electron captures of both K- and L-shells. If the initial and final states are degenerate in the context of radiative \ecec then a resonant enhancement in the decay rate could occur \cite{suj04}. The signal observed would come from the de-excitation photons.


The three combinations of electron capture and positron emission have not yet been observed, not even in the neutrino accompanied mode, though there has been a weak indication of the observation of $^{130}$Ba decay in geochemical experiments\cite{mes01}.

There is a wide variety of possible gamma ray emissions. This paper explores those double beta transitions of tin isotopes which emit gamma rays. There are three double beta isotopes of tin, \iso{Sn}{122} and \iso{Sn}{124} in the two electron mode and \iso{Sn}{112} for \bec and \ecec decays. The Q-values of the transition for each of the three isotopes are 366 keV, 2287 keV and 1922 keV and the natural abundances are 4.63\,\%, 5.79\,\% and 0.97\,\% respectively. As there is no excited state of interest for \iso{Sn}{122} decay we focus on the decays
\bea 
\iso{Sn}{124} &\ra& \tehv + 2e^- + (2 \bnel) + \gamma \\
2 e^- + \iso{Sn}{112} &\ra& \cdhz + (2 \nel) + \gamma \\
e^- + \iso{Sn}{112} &\ra& \cdhz + e^+ + (2 \nel) + \gamma 
\eea 
\\

\tinhv is one of the eleven isotopes with Q-values larger than 2\,MeV yet there is at present no proposal for a large scale experiment and few calculations.  There are no theoretical predictions of half-lives for 
$^{122}$Sn.  There are calculations reported for \zbb  ground state transitions in $^{124}$Sn, varying from $0.7-3 \times 10^{20}$\,yrs and estimates for the first excited states 0$^+$ and 2$^+$-states are $2.7\times 10^{21}$\,yrs and $6\times 10^{26}$\,yrs respectively.  \cite{aun96,cau99}


%

Calculations within the single state dominance model for the $2\nu$ \bec and $2\nu$ \ecec ground state transitions in \tinhz have been made\cite{dom05}. The half-lives expected are of the order of $10^{22}$\,yrs ($EC/EC$) and $10^{24}$\,yrs ($\beta^+/EC$). However it is possible that an enhancement of the rate for the \bec modes occurs due to right-handed weak currents.

The radiative neutrinoless \ecec rate to the ground state of \cdhz has been estimated to be of the order of $10^{-29}\,\mbox{yrs}^{-1}$.  This rate may be enhanced by a factor of $10^6$ by a possible resonance condition\cite{suj04}.  The characteristics of this mode are the monoenergetic internal bremsstrahlung gamma at $Q-E_K-E_L$ = 1888.5\,keV, with $E_K, E_L$ the K- and L-shell binding energy of $^{112}$Cd. There could also be a degeneracy between \tinhz and an excited 0$^+$ state at 1870.9\,keV in \mbox{$^{112}$Cd}, which would fulfil the resonance enhancement condition for neutrinoless \ecec\cite{bar07}.  

Two recent studies on tin decays produced half-life limits for various decay modes in the region of $10^{18}-10^{19}$\,yrs \cite{kim07,daw08}.  The aim of this measurement was to improve on those limits, if possible, by another order of magnitude. For this analysis, gamma line energies were taken from \cite{toi} and the latest atomic mass determinations from \cite{wap03}.

\section{Experimental setup}
The gamma-spectrometric measurements were performed in the Felsenkeller underground laboratory near Dresden (Germany) at a depth of 110\,mwe.  36 tin bars of 10.6\,cm$^2$ area and 4\,mm thickness plus 4 tin sheets of 24.2\,cm$^2 \times$  3\,mm were placed in a Marinelli beaker so that the endcap of the measuring detector was surrounded by the material. The total mass of tin used was 1.24\,kg. A passive shielding consisting of 170\,mm lead of graded qualities protected from the ambient background radiation in the measuring chamber. $^{222}$Rn concentration inside the shielding was reduced by flushing with gaseous nitrogen. The gamma spectrometer used a well-type low background p-type HPGe detector with 150\,cm$^3$ sensitive volume (30 \% relative efficiency, FWHM at 1.3\,MeV: 2.0\,keV). For further background suppression a U-shaped cryostat configuration was applied. Since the muon flux in the laboratory is reduced by a factor of 50 compared to the surface value, no additional active shielding was necessary. The pulse processing electronics used a standard spectrometric amplifier and the spectrum acquisition was performed by an 8\,k channel MCA. Spectra of the tin sample were accumulated during two measuring periods of 598 and 853\,hours duration, respectively. Thus, the total exposure amounts to 74.96 kg\,days (where the mass refers to natural tin). A blank spectrum of the empty shielding was measured for 762\,hours.\\
Just before data collection commenced, the Germanium detector was exposed to a mixed source of $^{241}$Am, $^{137}$Cs and $^{60}$Co. The gamma lines observed from this source were fitted with gaussian functions. The fitted means were used to accurately calibrate the ADC channel to energy relation and the widths were used to derive a function for the energy resolution (values given in Tab.~\ref{tab:exppara}). \\
The efficiency and response of the setup was studied using detailed GEANT4 based Monte Carlo simulations, which modelled all parts of the germanium crystal, the crystal housing, the tin pieces and the background shielding. Samples of $10^6$ gammas were simulated inside the tin volume for each energy of interest and the number of events in the photo-peak recorded in the crystal was used to determine the efficiency for observation, along with Poissonian uncertainties. Furthermore, know contaminants were simulated to confirm understanding of the shape of the background distribution. 

\section{Data}
\begin{figure}
\centering
         \includegraphics[width=\linewidth]{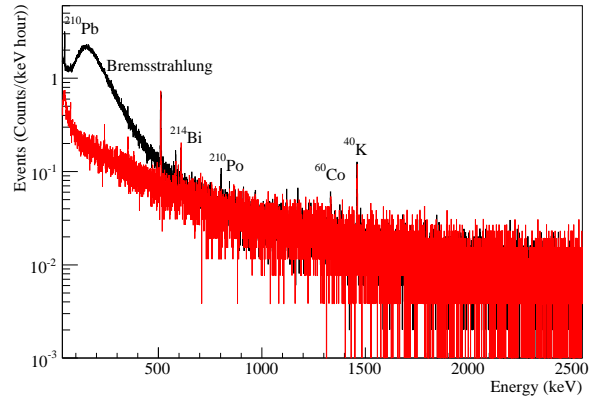}
\caption{Total spectrum (black), background spectrum (red). A clear bremsstrahlungs contribution at lower energies is seen.}
\label{comparedatback}
\end{figure}

\begin{figure}
\centering
\includegraphics[width=\linewidth]{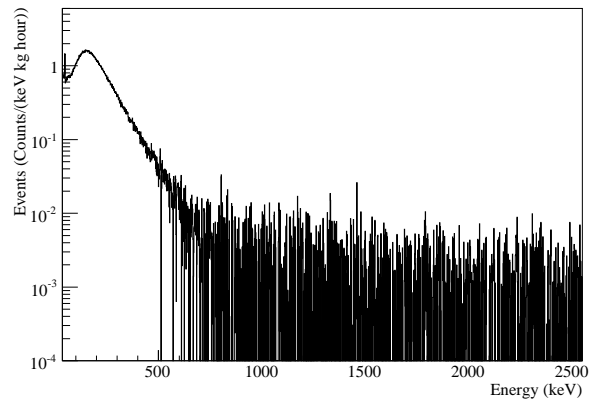}
\caption{Background subtracted and thus the spectrum of the tin sample.}
\label{subtracted}
\end{figure}

The obtained spectrum compared to a background run is shown in Fig.~\ref{comparedatback}. Due to the better low energy behaviour of the detector in the Felsenkeller with respect to the one used in \cite{daw08}  the 46.5\,keV line from $^{210}$Pb becomes visible, showing the presence of lead in the tin. $^{210}$Pb decays with a half life of 22.20 years via $^{210}$Bi (5.012\,days) and $^{210}$Po (138.38\,days) to stable $^{206}$Pb. Simulations were used to verify that the beta emitter $^{210}$Bi ($E_{max}$ = 1162.1 keV) is responsible for the bremstrahlungs background below 600\,keV. $^{210}$Po undergoes $\alpha$-decay but emits a weak gamma line at 803.10 keV which is visible in the tin spectrum. Further contamination of $^{214}$Pb and $^{214}$Bi ($^{222}$Rn  successors) as well as of $^{228}$Ra, $^{228}$Th, $^{40}$K, $^{137}$Cs and $^{60}$Co are visible in the tin spectrum at count rates below 0.2 counts/hr  (Fig~\ref{subtracted}).\par
Weak saw-tooth like features are observed in both tin and background spectra at 693\,keV and 596\,keV. These can be attributed to $^{72,74}$Ge(n,n') interactions~\cite{mei08}. The same features were observed more strongly in data collected in a surface laboratory, where the neutron background is expected to be higher, for Ref.~\cite{daw08}. These previous data were used to determine the functional form: 
\begin{eqnarray}
\label{e:neutron}
f(E) & = &\kappa.(1.0 - 0.04(E - E_0)) \\
& &~~when~~ (E>E_0)~\&~(1.0 - 0.04(E - E_0))>0 \nonumber\\
f(E) & = & 0~~	Otherwise \nonumber
\end{eqnarray} 
which was found to describe the shape of these features well. Here, $E_0$ is the starting point of the feature, taken to be 692\,keV and 595\,keV respectively, whilst $\kappa$ is the magnitude parameter, which varies with the neutron flux.

\section{Results}
The search performed relies on decays of \tinhv and $^{112}$Sn. The relevant experimental data, including energy resolution, efficiency and the background at the energy of each line searched for are compiled in Tab~\ref{tab:exppara}.

\begin{table}[htdp]
\begin{center}
\begin{tabular}{c|c|c|cc}
\hline\hline
Energy     (keV) & $\Delta E$ (keV)  & $\epsilon$ (\%)	  & \multicolumn{2}{|c}{B (counts/keV)}\\ 
		   &			   &		&  $\alpha$    & $\beta$ \\ \hline 
553.8	   &   1.739	   & 3.51		 & 476  & -0.62\\
602.7	   &   	1.763	   & 3.36		 & 425  & -0.49\\
606.5	   &   1.765	   & 3.38		 & 477  & -0.60\\
617.3	   &   1.770	   & 3.34		 & 533  & -0.69\\
694.7	   &   1.805	   & 3.06		 & 295  & -0.29\\
695.0	   &   1.806	   & 3.06		 & 303  & -0.30\\
713.8	   &   1.814	   & 3.01		 & 418  & -0.48\\
722.9	   &   1.818	   & 3.00		 & 377  & -0.42\\
815.1	   &   1.858	   & 2.76		 & 167  & -0.13\\
851.1	   &   1.873	   & 2.67		 & 139  & -0.10\\
1054.0         &   1.953	   & 2.31		 & 118  & -0.07\\
1253.4         &   2.022	   & 2.05		 & 126  & -0.07\\
1488.9         &   2.093	   & 1.82		 & 166  & -0.10\\
1888.5         &   2.184	   & 1.49		 & 123  & -0.06\\
\hline\hline
\end{tabular} 
\caption{\label{tab:exppara}The energy resolution, $\Delta E$ (FWHM), gamma detection efficiency, $\epsilon$, and fitted background, B, for each peak energy.
The magnitude of uncertainty in $\epsilon$ ranges from 0.01--0.02\%. 
The background continuum is defined by the polynomial $B = \alpha + \beta E$ and gives the total counts over the 1451\,hours of data collected. }
\end{center}
\end{table}

\subsection{The \tinhv system}
\begin{figure}
\centering
\includegraphics[width=\linewidth]{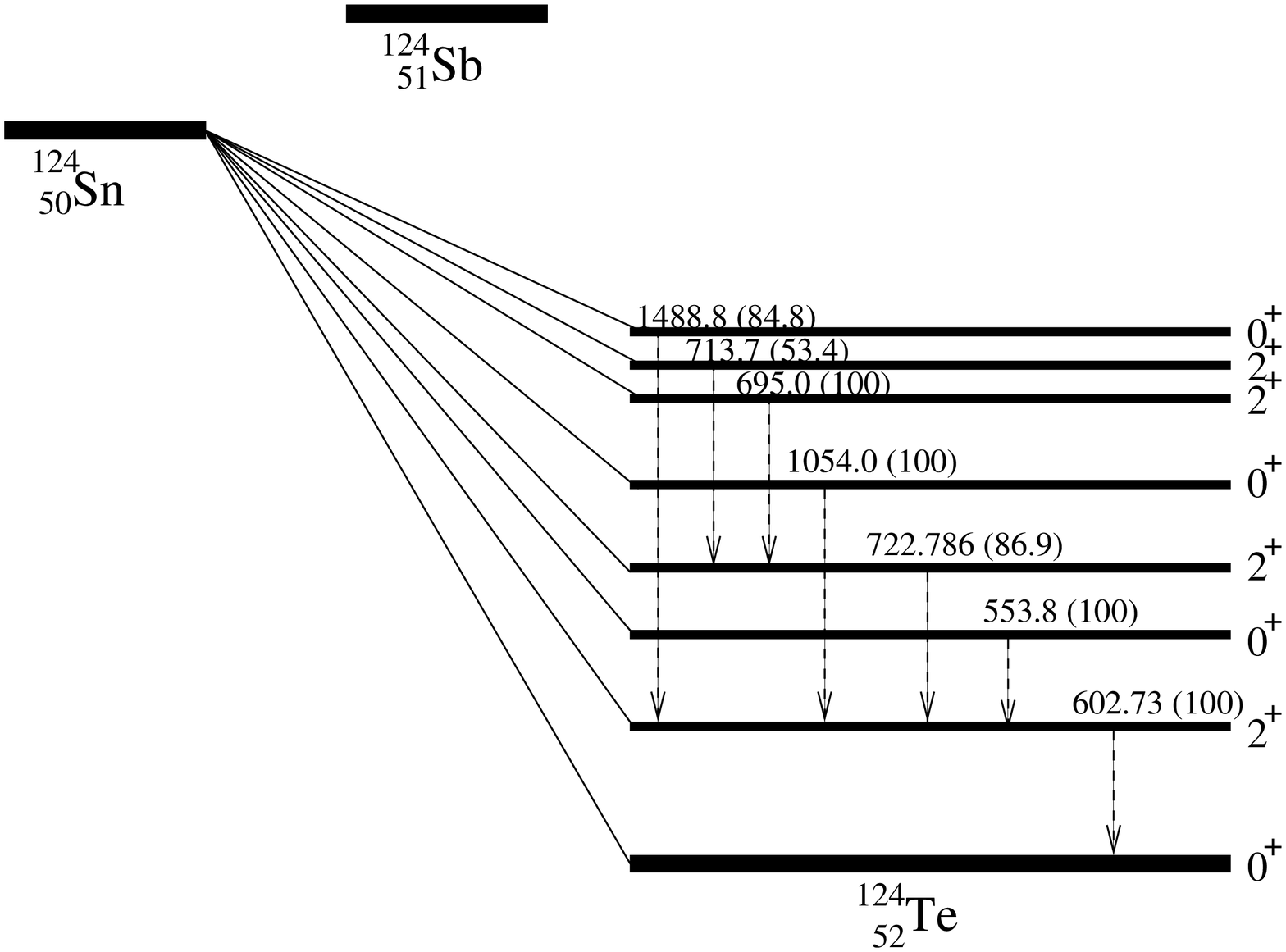}
\caption{Level diagram of $^{124}$Sn. The energy of the dominant mode decay gamma is given for each level along with the percentage branching ratio in brackets.}
\label{pic:levelhv}
\end{figure}

A Bayesian analysis approach~\cite{pdg-stats}, similar to that described in Ref.~\cite{daw08} was adopted. The most likely values for the magnitude of a linear background component and the amplitude of a gaussian peak at fixed gamma energy were determined from a binned maximum likelihood fit. The mean and width of the gamma-line under investigation were fixed in the fit, which was carried out over a range of $\pm 30 \sigma$ around the known gamma energy. When gamma lines from known radioactive contaminants fell within this fit range, their amplitude was also included as a fit parameter (again energy  and resolution were fixed). 
%
The effect of the $^{72,74}$Ge(n,n') interactions at 693\,keV and 596\,keV was also included, when relevant, by adding the function in Eqn.~\ref{e:neutron}, allowing the parameter $\kappa$ to vary as a parameter of the fit.\par
The fitted signal amplitude and uncertainty determined from the likelihood fit, $\theta_S \pm \delta_S$ was then used to derive a 90\% lower limit on the half-life for each decay mode using Eqn.~\ref{e:thalflim}.
\begin{eqnarray}
T_{\rm half} \ge \frac{N_{\rm iso} t_{\rm live} \epsilon \psi \ln2 }{\left(\theta_s +
1.28\delta_s\right)} (\theta_s > 0),\nonumber \\
T_{\rm half} \ge \frac{N_{\rm iso} t_{\rm live} \epsilon \psi \ln2 }{\left(
1.28\delta_s\right)} (\theta_s \leq 0)
\label{e:thalflim}
\end{eqnarray}
Here $N_{\rm iso}$ is the number of candidate nuclei in the tin for the given decay, $t_{\rm live}$ is the duration of data collection in years, $\epsilon$ is the efficiency for observing the given gamma signal determined from simulations and $\psi$ is the branching ratio for the gamma line in question. The factor of 1.28 gives one-sided 90\% limits. For each fit, the $\chi^2$ goodness of fit was determined and in all cases indicated good agreement with the data.

\begin{figure}
\centering
\includegraphics[width=\linewidth]{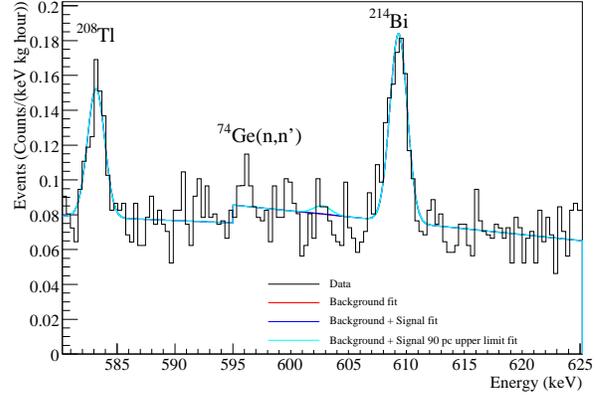}
\caption{Peak range around 602.7 keV. Natural radioactivity from $^{208}$Tl and $^{214}$Bi produces the lines at 583 and 609\,keV respectively. }
\label{pic:lineat603}
\end{figure}

The level scheme of \tehv (given in Fig.~\ref{pic:levelhv}) shows that the higher level decays all dominantly proceed via the intermediate $2_1^+$-state at 602.7\,keV. Hence, studies of a possible gamma peak at 602.7\,keV provide information on all these decays. The region of interest about this energy is shown in Fig.~\ref{pic:lineat603} along with the fitted signal gaussian and backgrounds. The efficiency for observing gammas of this energy from the tin was determined to be $3.36\pm$0.02\,\% from simulations and the calculated energy resolution at this point is 1.76\,keV (FWHM). With $1.4 \pm 4.6$ possible events in the gaussian peak a half-life of

\bea
\ton^{0(2)\nu} ( 0^+ \ra 2^+_1 (602.7\,keV)) &>& 8.7~(11.0) \times 10^{19}\mbox{yrs}\nonumber\\&~&(90~(68)~\% CL)
\eea
was obtained, an improvement of a factor 28 on existing limits~\cite{daw08} which can be attributed to the larger size of germanium detector and the reduced background due to the underground location of this experiment.

Although decays into higher excited states will produce two or more gammas, it is unlikely that they will all be emitted in the direction of the detector, and therefore, studies of higher states were performed by searching separately for the accompanying gammas. The fitted events and derived half-life limits for these decays are compiled in Tab.~\ref{tab:exstate-tinhv}. The branching ratio for the de-excitation chain, which is given in Fig.~\ref{pic:levelhv} and can be less than 100\% for the higher excited states, was included in the calculation of half-life. In some cases the best limit on half-life is derived from one of the lower de-excitation steps in the decay chain (indicated by a $\dag$ in the table).


%
\begin{table}[htbp]
\begin{center}
{\small
\begin{tabular}{|c|c|c|c|c|}
\hline
Excited state  & Gamma energy(ies)  & Events & \ton \\
& (keV) &  & ($10^{20}$ yrs)\\
\hline
$2^+_1$ (602.7) & {\bf602.7} & 0.6 $\pm$ 12.3 & 0.87\\
$0^+_1$ (1156.5) & {\bf553.8}, 602.7 & -44.4 $\pm$ 10.6 & 1.08 \\
$2^+_2$ (1325.5) & {\bf722.9}, 602.7 & 6.5$\pm$ 8.8 & 0.62$\dag$\\
$0^+_2$ (1656.7) & {\bf1054.0}, 602.7 & -1.8 $\pm$ 6.7 & 1.13 \\
$0^+_3$ (2020.0) & {\bf695.0}, 722.9, 602.7& 18.1  $\pm$12.1 & 0.38$\dag$ \\
$2^+_2$ (2039.3) & {\bf713.8}, 722.9, 602.7& -8.1 $\pm$ 8.5 & 0.62$\dag$ \\
$2^+_3$ (2091.6) & {\bf1488.9}, 602.7& -9.0 $\pm$ 4.7 & 1.08\\
\hline
\end{tabular}
}
\caption{\label{tab:exstate-tinhv}Half-life limits (90 \% CL) for all possible excited $0^+,2^+$-state transitions (0$\nu$ and 2$\nu$) of $^{124}$Sn.
Given are the excited states, all possible gamma lines with the dominant mode searched for in bold,  the total number of fitted events in the peak and deduced half-life limits. For decays marked \dag~the better half-life limit of the 602.7 keV gamma in the first row applies.}
\end{center}
\end{table}

\subsection{The \tinhz system}
\begin{figure}
\centering
\includegraphics[width=\linewidth]{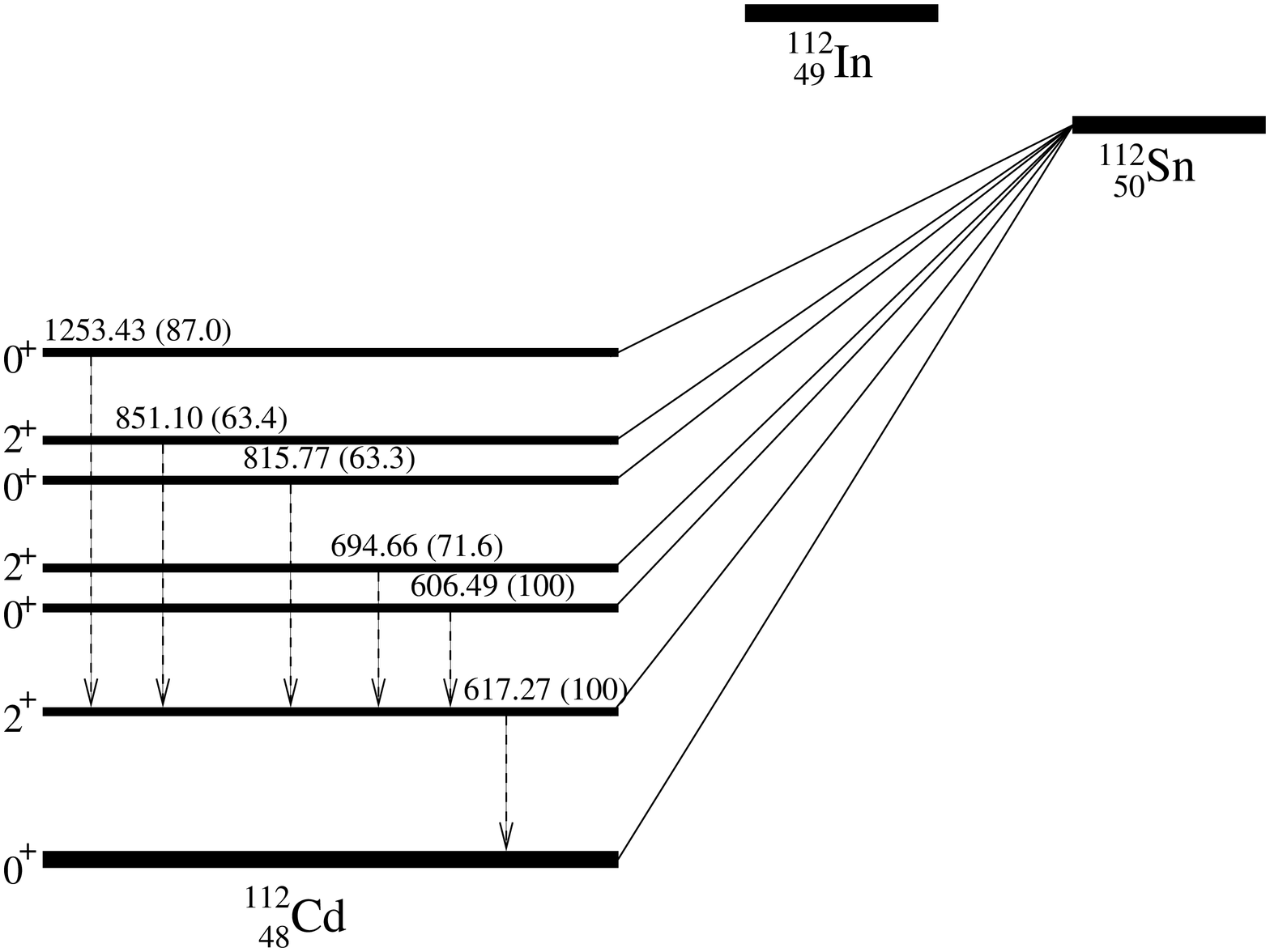}
\caption{Level diagram of $^{112}$Sn. The energy of the dominant mode decay gamma is given for each level along with the percentage branching ratio in brackets.}
\label{pic:levelhz}
\end{figure}

The second system to analyse is \tinhz, which has the decay level scheme shown in Fig.~\ref{pic:levelhz}. It offers a more complex search pattern as in addition to the de-excitation gammas more photons can be emitted in the \ecec or \bec process.

\begin{figure}
\centering
\includegraphics[width=\linewidth]{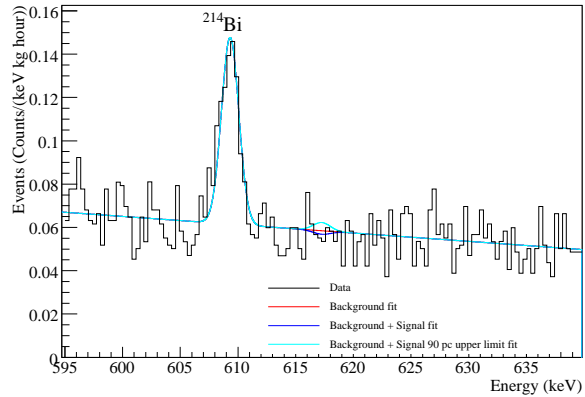}
\caption{Data in the range around 617.3\,keV. The line at 609.3\,keV stems from $^{214}$Bi  decay in the natural $^{238}$U decay chain.}
\label{pic:peak617}
\end{figure}

Again, all decays into higher excited states proceed via a $2_1^+$ state, this time at 617.3\,keV. Thus, a search for this line places a half-life limit on all excited state transitions, taking into account the branching ratios of the de-excitation. Fig.~\ref{pic:peak617} shows the fitted region where no peak was observed, resulting in a half-life limit of 
\bea
\label{eq:firstex}
\ton^{0(2)\nu EC/EC + \beta^+/EC} (0^+ \ra  2^+_1 (617.6 {\rm~keV}))>\nonumber\\ 1.8 \times 10^{19}~\mbox{yrs}~(90\% CL).
\eea
Decays to excited states were treated in a similar manner to the \tinhv system. The results of searches for all the potential gamma lines and the derived half-life limits are compiled in Tab.~\ref{tab:exstate-tinhz}.\par
Decay to the $0^+$-state at 1870.9\,keV is particularly interesting as this state could be degenerate with the ground state of \tinhz, which could result in a resonantly enhanced neutrinoless \ecec rate. De-excitation of this state is dominated by the emission of a 1253.4\,keV and 617.3\,keV gamma. No obvious peak was observed at either of these energies, resulting in a half-life limit on this decay mode of

\bea
\ton^{0(2)\nu EC/EC} ( 0^+ \ra 0^+(1870.9 {\rm~keV}))& &> 1.8 \times 10^{19} \mbox{yrs}\nonumber\\ &&(90\% CL).
\eea

The \bec modes can only populate excited states up to Q-1.022\,MeV, which corresponds to levels below 900\,keV for $^{112}$Sn. Only the $2^+_1$-state at 617.3\,keV satisfies this criteria. Therefore the half-life limit from the 617.6\,keV line search for \mbox{$EC/EC$}, given in Eqn.~\ref{eq:firstex} also applies for the 0(2)$\nu$ \bec-decay modes into the first excited $2^+_1$ state.  

\begin{table}[htbp]
\begin{center}
{\small
\begin{tabular}{|c|c|c|c|c|}
\hline
Excited state  & Gamma energy  & Events   & \ton \\
& (keV) & & ($10^{19}$ yrs)\\
\hline
$2^+_1$ (617.3) & {\bf617.6} & -5.4 $\pm$ 10.1 & 1.8\\
$0^+_1$ (1224.1) &  {\bf606.5}, 617.6  & -1.1 $\pm$ 10.4 & 1.8\\
$2^+_2$ (1312.3) & {\bf694.7}, 617.6 & 15.2 $\pm$ 12.1 & 0.5$\dag$\\
$0^+_2$ (1433.2) & {\bf815.8}, 617.6 & 8.3 $\pm$ 8.0 & 0.7$\dag$\\
$2^+_3$ (1468.7) & {\bf851.1}, 617.6 & -0.5 $\pm$ 7.4 & 1.3$\dag$ \\
$0^+_3$ (1870.9) & {\bf1253.4}, 617.6 & -5.0 $\pm$ 5.8 & 1.7$\dag$\\
\hline
\end{tabular}
}
\caption{\label{tab:exstate-tinhz}Half-life limits (90 \% CL) for excited state transitions (0$\nu$ and 2$\nu$) of $^{112}$Sn. Given are the excited states, all possible gamma lines with the dominant mode searched for in bold, the total number of fitted events in the peak and deduced half-life limits. For decays marked \dag~the better half-life limit of the 617.3 keV gamma in the first row applies.}
\end{center}
\end{table}

\par
\begin{figure}
\centering
\includegraphics[width=\linewidth]{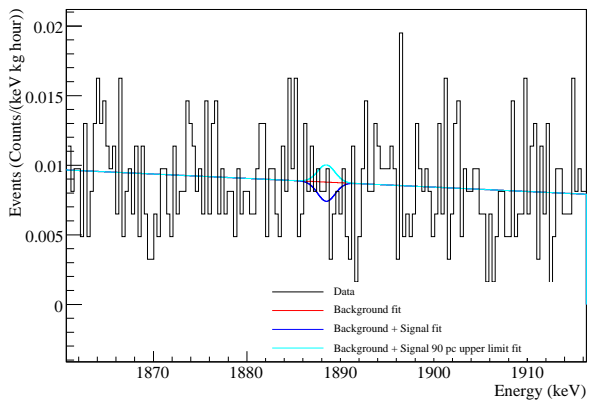}
\caption{Data around 1888.5 keV, and the fitted peak limits.}
\label{pic:peak1888}
\end{figure}

A search was also performed for the $0\nu$ \ecec ground state transition, based on the emission of an internal bremsstrahlung photon \cite{doi93}. Due to energy and momentum conservation this requires one K- and one L-shell capture. Additionally the bremsstrahlung photon has to be monoenergetic, with an energy of the Q-value reduced by the K- and L-shell binding energy of the daughter. In the case of \tinhz this implies a gamma of 1888.5\,keV. The calculated detection efficiency for such a photon is 1.491 $\pm$ 0.013\,\%. Fig.~\ref{pic:peak1888} shows the energy region for this search, which resulted in a half-life limit of 

\be
T_{1/2}>2.0\times 10^{19} \mbox{yrs}~(90\% CL)
\ee
due to the non-observation of a peak. \\
Uncertainties in the atomic masses mean that the peak position is only know to about $\pm2$\,keV~\cite{wap03}. Therefore,  the peak position was systematically varied by 0.5\,keV between 1886 and 1890\,keV and the fit was repeated for the corresponding position. The conservative lower limit for this decay is taken as 
\be
\ton^{0\nu EC/EC} ( 0^+ \ra 0^+_{g.s}) > 0.8 \times 10^{19} \mbox{yrs}~(90\% CL),
\ee
the worst half-life limit found in this range.
 
\section{Summary and Conclusions}
Double beta decay, in its various forms, is a powerful tool for investigating neutrino properties. Although it has received little interest in the field so far, tin presents two isotopes of interest:
\tinhv which is suitable for $0\nu\beta^-\beta^-$ searches and \tinhz available for decay via the \bec mode. In this paper, new half-life limits are presented for various decays of these two isotopes, based on the non-observation of characteristic gamma-lines. All of the limits presented here improve on previous values given in Ref.~\cite{daw08} by an order of magnitude or more. With limits of order 10$^{20}$\,yrs for a number of decay modes, this paper might trigger some new theoretical efforts to calculate \zbb half-lives in the tin systems. During submission of this paper we became aware of a similar study in Ref.~\cite{bar08}.

\section{Acknowledgement}
We thank V.I. Tretyak for useful discussion, and B. Morgan and Y. Ramachers for software. J.R. Wilson acknowledges support from the Leverhulme Trust.

\end{document}